# Low-temperature quantum transport in CVD-grown single crystal graphene

Shaohua Xiang,[1] Vaidotas Miseikis,[2] Luca Planat,[1] Stefano Guiducci,[1] Stefano Roddaro,[1] Camilla Coletti,[2] Fabio Beltram,[1,2] and Stefan Heun[1](✉)

[1]NEST, Istituto Nanoscienze—CNR and Scuola Normale Superiore, Piazza San Silvestro 12, 56127 Pisa, Italy
[2]Center for Nanotechnology Innovation @NEST, Istituto Italiano di Tecnologia, Piazza San Silvestro 12, 56127 Pisa, Italy

## ABSTRACT

Chemical vapor deposition (CVD) has been proposed for large-scale graphene synthesis for practical applications. However, the inferior electronic properties of CVD graphene are one of the key problems to be solved. In this study, we present a detailed study on the electronic properties of high-quality single crystal monolayer graphene. The graphene is grown by CVD on copper using a cold-wall reactor and then transferred to Si/SiO$_2$. Our low-temperature magneto-transport data demonstrate that the characteristics of the measured single-crystal CVD graphene samples are superior to those of polycrystalline graphene and have a quality which is comparable to that of exfoliated graphene on Si/SiO$_2$. The Dirac point in our best samples is located at back-gate voltages of less than 10V, and their mobility can reach 11000 cm$^2$/Vs. More than 12 flat and discernible half-integer quantum Hall plateaus have been observed in high magnetic field on both the electron and hole side of the Dirac point. At low magnetic field, the magnetoresistance shows a clear weak localization peak. Using the theory of McCann *et al.*, we find that the inelastic scattering length is larger than 1 μm in these samples even at the charge neutrality point.

## 1. Introduction

Graphene is known to be a very attractive and promising material for the microelectronics industry. Mechanically exfoliated graphene [1,2] still remains the preferred way to produce graphene for fundamental investigations since this is the simplest way to obtain high-quality graphene flakes. Yet exfoliation is not a scalable method and cannot yield large graphene samples of interest for practical applications. Various methods were proposed for large-scale graphene synthesis, such as epitaxial growth on SiC surfaces [3-5] and chemical vapor deposition (CVD) on transition metals [6-13]. CVD is a simple and low-cost method for large-scale graphene synthesis, and it is easy to transfer the grown graphene to other substrates. Unfortunately the quality of most CVD graphene is not yet comparable to that of exfoliated graphene, because CVD graphene is typically polycrystalline, and the lattice mismatch between graphene and the metal substrate introduces defects in CVD graphene [14].

Several protocols were reported to improve the electrical characteristics of CVD-graphene, including a dry transfer technique and the use of flakes of hexagonal boron nitride (h-BN) as substrates [15]. Dry transfer can reduce the transfer-related contamination thus minimizing the extrinsic doping of CVD graphene. Furthermore, h-BN substrates can provide atomically-smooth surfaces and small lattice mismatch, which leads to reduced substrate-induced scattering. Unfortunately, owing to the small size of h-BN flakes, this method requires a cumbersome aligned-transfer of graphene, and device dimensions are limited [15]. With an aim to enable large-scale devices of high-quality graphene, we recently presented a method for CVD synthesis of large single crystals of graphene, up to several millimeters in size [12]. This technique produces well-isolated large crystals of graphene, which can be readily contacted for electrical characterization without the need of graphene etching, avoiding the consequent potential degradation of device performance. Furthermore, in the present work we transfer graphene on Si/SiO$_2$ by employing electrochemical delamination to remove the graphene from the growth substrate [16] instead of the traditional method, which relies on wet-etching of copper [9]. This prevents unintentional doping of graphene by the etchant residues, a common issue with transferred CVD-graphene graphene [17].

In our previous work, CVD graphene was characterized by optical microscopy, scanning electron microscopy, x-ray photoelectron spectroscopy, Raman spectroscopy as well as selected area-electron diffraction and low-energy electron diffraction [12]. Here, we focus on investigating quantum-Hall effect (QHE) and weak localization (WL) that are often used to evaluate the quality of graphene sheets as probes of carrier scattering at low temperatures. The QHE in graphene is different from other conventional 2D systems, owing to the existence of a non-zero Berry phase in the electron wave function [18,19]. This leads to the formation of half-integer quantum Hall plateaus that were observed in exfoliated [1,2,20,21] and CVD graphene [15,22-24]. Compared to these latter data on CVD graphene, and even to data reported for single-crystal CVD graphene deposited on h-BN [15], here we demonstrate much improved QHE data: we observe more than 12 flat and discernible half-integer quantum Hall plateaus. These results are comparable to the best observations reported for exfoliated graphene on SiO$_2$. The WL (or weak anti-localization (WAL)) is one of the most studied effects in graphene [22-23,25-34]. WL is a powerful tool to probe the carrier scattering and phase coherence in graphene. In this study, we show inelastic scattering lengths exceeding 1 μm even at the Dirac point, which favorably compares to the performance of exfoliated graphene. Analyzing the dephasing scattering time, we find that the electron-electron interaction is the main inelastic scattering mechanism in our samples.

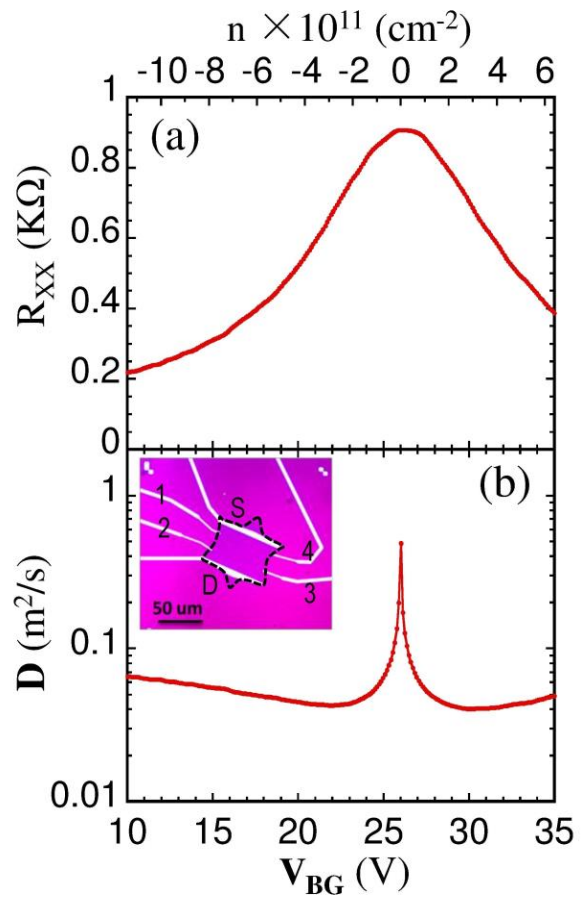

**Figure 1**: (a) Four-point longitudinal resistance as a function of back-gate voltage $V_{BG}$. (b) Diffusion coefficient as a function of back-gate voltage $V_{BG}$. $B = 0$ T, $T = 250$ mK, Device-1. The upper $x$-axis shows the back-gate voltage converted to carrier concentration $n$. The inset shows an optical microscopy image of Device-1.

2. **Experimental**

Large crystals of CVD graphene were synthesized on a copper foil by low-pressure CVD as described in Ref. [12]. Samples were spin-coated with a poly(methyl methacrylate) (PMMA) support film, and the graphene/PMMA stack was removed from the Cu substrate by electrochemical delamination [16]. It was then transferred onto 300 nm SiO$_2$ on degenerately doped Si substrates. The devices investigated in this study are as large as 50 μm × 70 μm. Metallic contacts (Cr/Au: 5 nm/60 nm) were defined by electron beam lithography and thermal evaporation. All data reported in this article were obtained from two devices. The optical microscopy image of Device-1 is shown in the inset of Fig. 1(b) (Device-2 has the same configuration). In order to tune carrier density in these graphene devices, a back-gate voltage ($V_{BG}$) was applied during measurements. All measurements were done using a lock-in technique in a Heliox Helium-3 cryostat with base temperature of 250 mK. A four-point configuration was used for magnetoresistance measurement. The longitudinal and transversal resistances are defined as $R_{xx} = V_{12}/I_{SD}$ and $R_{xy} = V_{14}/I_{SD}$, respectively, with $V_{ij}$ the voltage drop measured between contacts $i$ and $j$ and $I_{SD}$ the applied source-drain current.

## 3. Results and Discussion

Figure 1(a) shows the measured four-point longitudinal resistance $R_{xx}$ as a function of back-gate voltage $V_{BG}$ for Device-1 at zero magnetic field, with a peak at $V_{BG} \approx 26$ V (the Dirac point of Device-1). The same measurement was also performed on Device-2, and the Dirac point was observed at around 9 V (see the inset of Fig. 2(a)). The upper $x$-axis of Fig. 1 shows the back-gate voltage converted to carrier concentration $n \approx C_G |V_{BG} - V_{Dirac}|/e$ [35], where $C_G$ is the gate capacitance per area (11.5 nF/cm$^2$, for 300 nm oxide) and $e$ is the elementary charge. The mobility $\mu$ of our devices was obtained using the formula $\mu = 1/ne\rho$, with $\rho$ the resistivity at $B = 0$ T. The four point mobility away from the Dirac point is 8200 cm$^2$V$^{-1}$s$^{-1}$ for Device-1, while for Device-2 we obtained 11000 cm$^2$V$^{-1}$s$^{-1}$. Figure 1(b) shows the diffusion coefficient $D$ as a function of back-gate voltage. The diffusion coefficient $D = v_F l / 2$ is determined from the mean free path $l = h/2e^2 k_F \rho$ [28], where $v_F$ is the Fermi velocity (1.1×10$^6$ m/s for graphene [35]), $k_F = (\pi n)^{1/2}$ is the value of the Fermi wave number. This simple model does not consider a residual carrier density at the Dirac point due to the formation of electron-hole puddles and will therefore overestimate $D$ close to the Dirac point and lead to a cusp-like artifact.

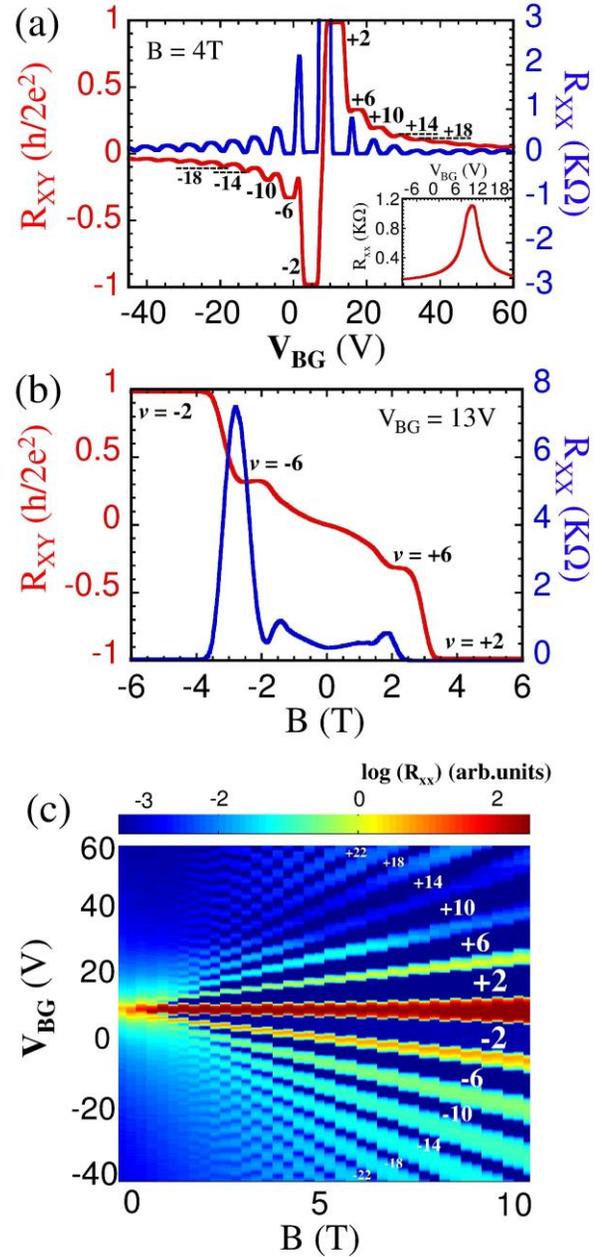

**Figure 2**: (a) Hall resistance $R_{xy}$ (red, left) and longitudinal resistance $R_{xx}$ (blue, right) as a function of $V_{BG}$ at $B = 4$ T, $T = 250$ mK for Device-2. The inset shows the four point longitudinal resistance of Device-2 as a function of back-gate voltage $V_{BG}$ at $B = 0$ T, $T = 250$ mK. (b) The Hall resistance $R_{xy}$ (red, left) and longitudinal resistance $R_{xx}$ (blue, right) as a function of magnetic field $B$ at $V_{BG} = 13$ V, $T = 250$ mK. (c) Landau fan diagram of longitudinal resistance $R_{xx}$ as a function of $V_{BG}$ and $B$ at $T = 250$ mK for Device-2.

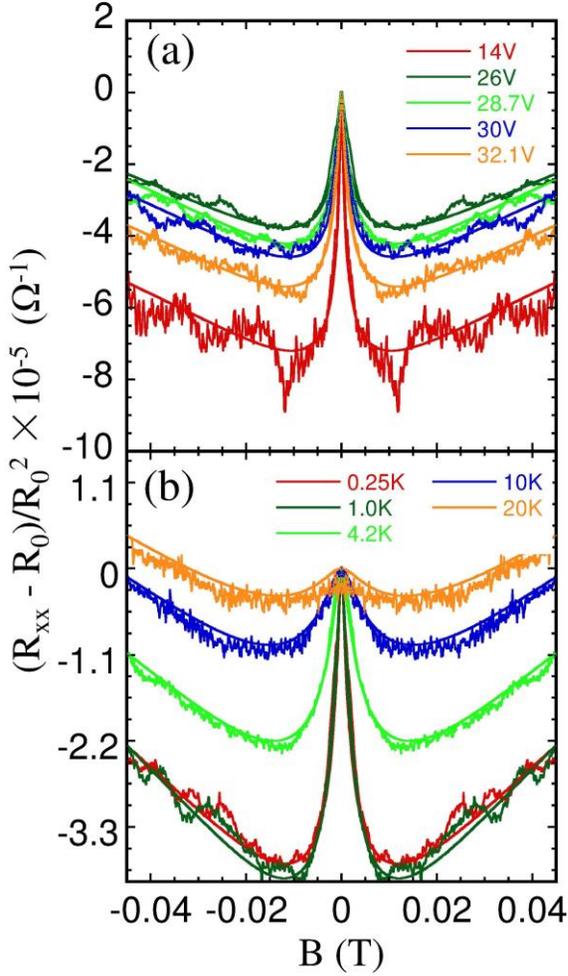

**Figure 3:** (a) Normalized magnetoresistance $(R_{xx} - R_0)/R_0^2$ of Device-1 as a function of magnetic field $B$ for different back-gate voltages. (b) Normalized magnetoresistance as a function of magnetic field $B$ for different temperatures. The solid curves in (a) and (b) are best fits to Eq. (1).

Figure 2(a) shows the quantum Hall effect of Device-2 at 250 mK. The magnetic field was set at 4.0 T and the back-gate voltage was swept from -40 V to 60 V. Several half-integer quantum Hall plateaus are observed, both for the electron and hole branches. The QHE trace recorded at fixed back-gate voltage (13 V) and sweeping the magnetic field is shown in Fig. 2(b). The $\nu = \pm 6$ plateaus are centered at ±2.2 T, which indicates a carrier density of $(3.2 \pm 0.1) \times 10^{11}$ cm$^{-2}$. This is consistent with the value extracted from the $R_{xx}(V_{BG})$ curve at $V_{BG}$ = 13 V (inset to Fig. 2(a)) which is $(2.9 \pm 0.2) \times 10^{11}$ cm$^{-2}$. In general, the intrinsic carrier concentration in this sample is low (hole concentration $6.4 \times 10^{11}$ cm$^{-2}$ for $V_{BG}$ = 0 V). Indeed it is one order of magnitude lower than values typically reported for CVD graphene on SiO$_2$ [22,23,32,34] and indicates the low level of impurities and defects in our material. Samples of comparable quality have been reported for exfoliated graphene on SiO$_2$, which are typically considered to contain a low level of defects. This suggests that the SiO$_2$ substrate is a key limiting factor in the studied devices [36].

Figure 2(c) shows the Landau fan diagram for Device-2, with log($R_{xx}$) plotted as a function of magnetic field $B$ and back-gate voltage $V_{BG}$: more than 12 Landau levels are clearly observed. Compared with previous reports on CVD graphene [22,23], the QHE here is much improved with respect to devices implemented on Si/SiO$_2$.

Weak localization is resulting from quantum interference effects in 2D systems [37] that can lead to an enhanced carrier back-scattering probability, i.e. to an increased zero-field resistance. Notably, resistance will recover its classical value at finite magnetic fields, since the interference effect is suppressed at even moderate non-zero fields. Figure 3(a) shows the normalized [33] and symmetrized [34] magnetoresistance data $(R_{xx} - R_0)/R_0^2$, with $R_0$ the resistance at $B$ = 0 T. Data were taken from Device-1 at 250 mK for several selected back-gate voltages. All curves show the conventional character of WL with a resistivity peak around $B$ = 0 T. The temperature dependence of the WL at the Dirac point ($V_{BG}$ = 26 V) is shown in Fig. 3(b). It shows the behavior already observed in exfoliated graphene [27,28]. The WL peak broadens with increasing temperature and vanishes at higher temperatures (for Device-2, the peak completely disappears at 20 K). At the lowest temperatures, weak oscillations are visible in the magnetoresistance data, superimposed to the WL traces. These are most likely mesoscopic conductance fluctuations as suggested by their disappearing as temperature is increased (see Fig. 3(b) and Ref. [38]).

The phenomenology of WL in graphene is different from that observed in conventional 2D systems. It is not only sensitive to inelastic scattering and to spin-flip processes, but also to breaking of the chiral symmetry [39-41]. For the analysis of the scattering in our CVD graphene samples, we refer to the theory of McCann *et al.* [40] and the function used in Ref. [33]:

$$\frac{\Delta R_{xx}}{R_0^2} = -\frac{e^2}{\pi h}\left[F\left(\frac{\tau_B^{-1}}{\tau_\varphi^{-1}}\right) - F\left(\frac{\tau_B^{-1}}{\tau_\varphi^{-1}+2\tau_{iv}^{-1}}\right) - 2F\left(\frac{\tau_B^{-1}}{\tau_\varphi^{-1}+\tau_*^{-1}}\right)\right],$$

(1)

with $\Delta R_{xx} = R_{xx} - R_0$, while $\tau_\varphi$, $\tau_{iv}$, and $\tau_*$ are the inelastic dephasing (or decoherence) time, (elastic) intervalley scattering time, and (elastic) intravalley scattering time, respectively. $F(z) = ln(z) + \psi(0.5 + z^{-1})$, and $\psi(x)$ is the digamma function. $\tau_B^{-1} = 4DeB/\hbar$, with $D$ the diffusion coefficient. The solid curves in Fig. 3 are the best fits of the data to Eq. (1). From these fits, the values of $\tau_\varphi$, $\tau_{iv}$, and $\tau_*$ can be obtained for different back-gate voltages and temperatures.

Figure 4(a) shows the fitting results for different back-gate voltages at a temperature of 250 mK. Since $\tau_B \propto D^{-1}$, artefacts related to the divergence of $D$ at the Dirac point are transferred to the extracted values of the characteristic scattering times. This can be avoided by considering the scattering length $L = (D\tau)^{1/2}$. The results are shown in Figs. 4(b) and (c). The intervalley scattering length $L_{iv}$ and the intravalley scattering length $L_*$ are almost carrier-density independent. On the other hand, the dephasing scattering length $L_\varphi$ is dependent on carrier-density. Its lowest value, at the Dirac point, is as high as 1 µm, and even higher values are observed with increasing electron (hole) density. In other words, the inelastic scattering is strongly dependent on carrier density, while the elastic scattering is almost carrier-density independent. We also find that the intervalley scattering length $L_{iv}$ is smaller than $L_\varphi$, while the intravalley scattering length $L_*$ is smaller than $L_{iv}$ and $L_\varphi$ ($L_\varphi > L_{iv} > L_*$). This behavior is similar to previous observations in graphene [22,29,31,32]. In particular, in Ref. [29], the minimum value of $L_\varphi$ at the Dirac point was attributed to a charge inhomogeneity of the sample at low carrier density that results in the formation of electron-hole puddles.

The temperature dependence of the scattering lengths for three selected back-gate voltages is shown in Fig. 4(c). We found that the intervalley scattering length $L_{iv}$ and the intravalley scattering length $L_*$ are only weakly temperature dependent. On the other hand, the dephasing scattering length $L_\varphi$ is not only carrier-density dependent, but also strongly temperature dependent. $L_\varphi$ increases with lower temperatures and saturates for the lowest temperatures at a temperature $T_{sat}$. We note that this cannot be attributed to a saturation of the electron temperature in the sample, because in previous experiments on similar samples we have observed clear differences between measurements performed at 1 K and at 250 mK [33]. A saturation behavior of $L_\varphi$ was also pointed out in other reports [27-29], however, all of them were observed in exfoliated graphene. In Refs. [28] and [29] it was demonstrated that $T_{sat}$ is also carrier-density dependent. D. K. Ki *et al.* suggested that the observed saturation of $L_\varphi$ at the lowest temperatures could be related to electron-hole puddles [29]. It is worth noting that the dephasing length $L_\varphi$ found in this study is much larger than the results reported in previous studies on CVD graphene [22,23,32]. Not only does this demonstrate the high quality of our CVD graphene, but it also explains why in these previous studies the dephasing length did not reach saturation at the lowest measurement temperatures.

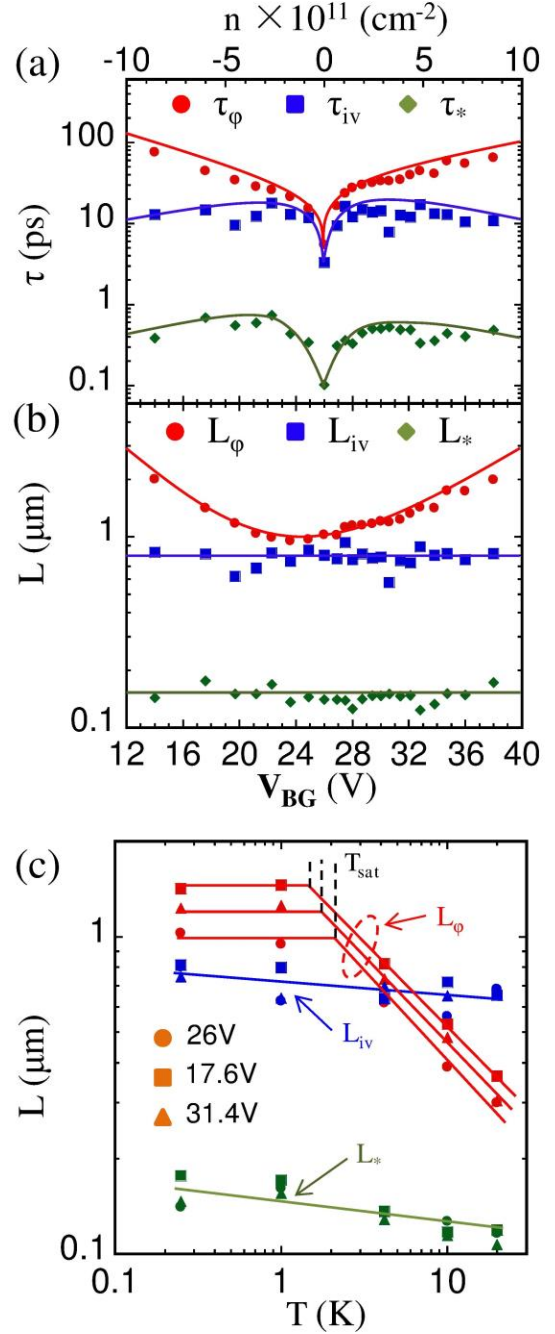

**Figure 4:** (a) Scattering times as a function of back-gate voltage $V_{BG}$ at $T = 250$ mK. The solid lines are guides to the eye. (b) Scattering lengths as a function of back-gate voltage $V_{BG}$ at $T = 250$ mK. The solid lines are guides to the eye. The upper *x*-axis shows the back-gate voltage converted to carrier concentration *n*. (c) Scattering lengths as a function of temperature *T* for three selected back-gate voltages. The red, blue, and green markers correspond to dephasing length, intervalley scattering length, and intravalley scattering length, respectively. The solid lines are guides to the eye. All results are obtained from Device-1.

At low temperature, the dominant mechanism of inelastic scattering in graphene is considered to be the electron-electron interaction [27-29,42]. This is consistent with our observation that only the inelastic scattering time is strongly dependent on carrier density (see Fig. 4(a)). To examine this further, we analyzed the temperature dependence of the dephasing time found by fitting the experimental data to Eq. (1). The dephasing time $\tau_\varphi$ as a function of $T^{-1}$ is shown in Fig. 5(a) for three selected back-gate voltages (i.e. carrier densities). A saturation of the inelastic scattering time is observed at around 1 K$^{-1}$. Such saturation was already observed in exfoliated and epitaxial graphene [27,28,33,42,43]. Before saturation, $\tau_\varphi$ shows a linear dependence on $T^{-1}$ which is in agreement with previous reports [27,28]. This is clear evidence of the fact that electron-electron interaction is the main inelastic-scattering mechanism in our CVD graphene. Figure 5(b) shows the dephasing scattering rate as a function of dimensionless conductivity $g(n) = \sigma(n)h/e^2$ at 250 mK, where the $\sigma(n)$ is conductivity, $n$ is carrier density, and $h$ is Planck's constant. The red dashed line is plotted according to the Nyquist scattering function [43,44] described as [29]:

$$\frac{1}{\tau_\varphi} \propto a k_B T \frac{\ln(g)}{\hbar g} \quad . \tag{2}$$

Figure 5(b) shows that Eq. (2), with $a$~6.5, nicely describes $\tau_\varphi^{-1}$ values for g(n) > 10, i.e. for back-gate voltage beyond about 4 V from the Dirac point. Closer to the Dirac point Eq. (2) predicts $\tau_\varphi^{-1}$ values which are clearly too small and fail to reproduce the observed experimental behavior. This result is similar what is reported in Ref. [29] where authors suggested that the main inelastic scattering mechanism in graphene is electron-electron interaction; however, near the Dirac point, an additional inelastic scattering mechanism becomes dominant that likely is caused by electron-hole puddles.

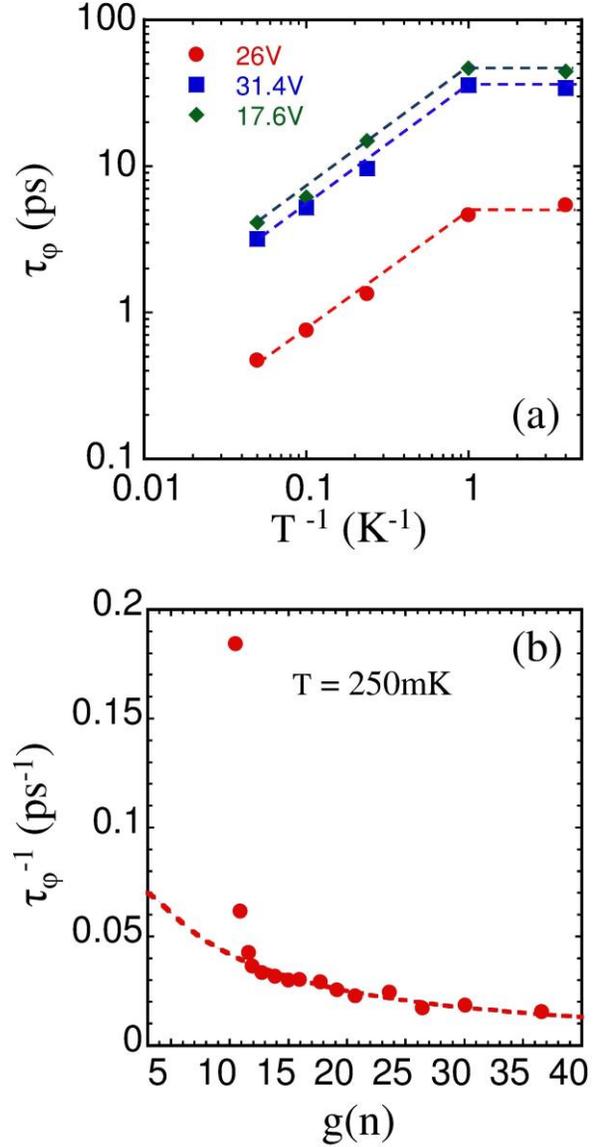

**Figure 5**: (a) Dephasing time as a function of inverse temperature $1/T$ for three selected back-gate voltages. The dashed lines are a guide to the eye. (b) Dephasing scattering rate as a function of dimensionless conductance $g$ at $T$ = 250 mK. The dashed curve is the best fit to Eq. (2). All results are obtained from Device-1.

## 4. Conclusion

We measured the magnetoresistance and quantum interference effects in single-crystal CVD graphene at low temperatures. Flat and discernible half-integer quantum Hall plateaus were observed up to filling factor ν = ± 30 by sweeping back-gate voltage at a fixed magnetic field, or by sweeping the magnetic field at a fixed back-gate voltage. In order to investigate the scattering mechanisms in our CVD graphene, we measured the magnetoresistance at low magnetic field for several back-gate voltages and several temperatures. Weak localization was observed even at the Dirac point. We fitted our WL data by using the theory of E. McCann *et al*. The results were similar to observations in exfoliated graphene. Furthermore, we found that the measured dephasing length is much larger than that observed in other groups' polycrystalline CVD graphene. As in exfoliated graphene, the electron-electron interaction is the main dephasing scattering mechanism. In summary, these measurements demonstrate that by CVD it is possible to obtain graphene devices on $SiO_2$ having an electronic quality comparable to those obtained by exfoliation.

## Acknowledgements

The authors acknowledge financial support from the Italian Ministry of Foreign Affairs (Ministero degli Affari Esteri, Direzione Generale per la Promozione del Sistema Paese) in the framework of the agreement on scientific collaborations with Canada (Quebec) and Poland; and from the CNR in the framework of the agreement on scientific collaborations between CNR and JSPS (Japan), CNRS (France), and RFBR (Russia). We also acknowledge funding from the European Union Seventh Framework Programme under grant agreement no. 604391 Graphene Flagship. S. G. acknowledges support by Fondazione Silvio Tronchetti Provera.